\documentclass[prd,onecolumn]{revtex4}
\usepackage{graphicx}
\usepackage{amssymb}
\usepackage{epstopdf}

\begin{document}

\title{Special and General Relativistic Effects in Galactic Rotation Curves}
\date{\today}

\author{Alan Cooney, Dimitrios Psaltis, Dennis Zaritsky}
\affiliation{Steward Observatory and Department of Physics, 
University of Arizona, 933 N.\ Cherry Ave, Tucson, AZ 85721, USA}

\begin{abstract}
The observed flat rotation curves of galaxies require either the
presence of dark matter in Newtonian gravitational potentials or a
significant modification to the theory of gravity at galactic
scales. Detecting relativistic Doppler shifts and gravitational
effects in the rotation curves offers a tool for distinguishing
between predictions of gravity theories that modify the inertia of
particles and those that modify the field equations. These higher-order 
effects also allow us in principle, to test whether dark
matter particles obey the equivalence principle. We calculate here the
magnitudes of the relativistic Doppler and gravitational shifts
expected in realistic models of galaxies in a general metric theory of
gravity. We identify a number of observable quantities that measure
independently the special- and general-relativistic effects in each
galaxy and suggest that both effects might be detected in a statistical
sense by combining appropriately the rotation curves of a large number
of galaxies.
\end{abstract}

\maketitle 

\section{Introduction}

The rotation curves of galaxies are direct probes of the shape of
their gravitational fields and their matter content.  That the
inferred circular velocities remain approximately constant, even at
large distances from the central luminous matter, is the strongest
evidence for the presence of dark matter at galactic
scales~\cite{dmhalos}.

Despite its simplicity and remarkable success in accounting for
observations over a wide range of scales, explaining galactic rotation
curves with dark matter remains a hypothesis; to date the candidate
dark matter particles have eluded direct detection~\cite{dm}. In the
meantime, several possible modifications to the theory of gravity have
been explored in attempts to explain the observed galactic rotation
curves The MOdified Newtonian Dynamics (MOND) framework has been the
most successful attempt phenomenologically~\cite{MOND,MONDrev}, but
suffers from the fact that it is not relativistic. As a result, it
cannot be used in its empirical form to generate predictions for
gravitational lensing or for the dynamical evolution of systems at
scales comparable to the Hubble scale.

Earlier~\cite{relMOND} attempts to develop a relativistic theory of
gravity that mimics the MOND phenomenology often faced fundamental
difficulties, such as problems with causality or inadequacies in
accounting for gravitational lensing in galaxies. More recently, new
models~\cite{relMONDnew} have been developed to resolve these
difficulties, but they introduce several additional fields and
auxiliary functions. Such additions negate the most appealing aspect
of the original MOND, i.e., that galactic rotation curves and the
Tully-Fisher relation were accounted for with the introduction of a
single acceleration scale.

Most previous attempts aim to reproduce the MOND phenomenology by
modifying the general relativistic field equations. However, in
principle, the MOND phenomenology can also be achieved in the
non-relativistic limit by modifying the equivalence principle, i.e.,
the inertia of test particles~\cite{inertia}.Which of the two aspects
of the theory of gravity need to be altered in order to account for
the observed rotation curves of galaxies, in the absence of dark
matter? This question cannot be resolved solely with observations of
the non-relativistic Doppler shifts of tracer particles. 
Ideas involving violations of the equivalence principle have been tested
empirically, within the dark matter interpretation of galactic
rotation curves~\cite{equiv}, but not in the framework of modified
gravity

In this article, we aim to address this question by calculating the
second order special- and general-relativistic corrections to the
Doppler shifts of atomic lines used to infer the rotation curves of
galaxies. Similar calculations in General Relativity, as well as
strategies for using such measurements to map the spacetimes of
galaxies have been reported in previous studies~\cite{rot}. Here we
focus only on the second-order effects and evaluate them using only
the symmetries of the spacetime. More importantly, we show explicitly
that, if the underlying theory of gravity obeys the equivalence
principle, then the second-order effects can be determined entirely
using knowledge of the non-relativistic Doppler shifts and without any
assumptions regarding the underlying field equations of the theory.
As a result, the relation between non-relativistic and relativistic
effects can be used as a test of the equivalence principle at galactic
scales, independent of whether dark matter or modified gravitational
field equations are responsible for the flat rotation curves.

\section{Particles And Photons In Galactic Potentials}
\subsection{The Circular Orbits of Particles}

We begin by assuming that the gravitational potential of a galaxy
exhibits a high degree of axisymmetry.  We will ultimately calculate
the second-order Doppler shifts and the gravitational corrections
assuming that the spacetime of each galaxy is asymptotically
flat. This approach is formally appropriate only for a galaxy in an
otherwise empty Universe; in a subsequent section, we correct for the
Cosmological redshift of the galaxy.

Our assumptions motivate our choice of metric
\begin{equation}
ds^2 = g_{tt}\left(r,\theta\right)  dt^2 + 2g_{t\phi}\left(r,\theta\right) \, dt\, d\phi  + g_{rr}\left(r,\theta\right) dr^2 +g_{\theta \theta}\left(r,\theta\right)  d\theta^2 + r^2 \sin^2 \theta \,d\phi^2\;,
\label{eq:metric}
\end{equation}
where $g_{tt}$, $g_{t\phi}$, $g_{rr}$, and $g_{\theta \theta}$ are
undetermined functions of the coordinate radius $r$ and the polar
angle $\theta$. The $g_{t\phi}$ coefficient is associated with
frame-dragging, which we expect not to be significant for galaxies and
will be neglected henceforth. For comparison, we recall that in
General Relativity, the external spacetime of a spherically symmetric
object is unique and is given by the Schwarzschild solution, for which
$g_{t\phi} =0$, $g_{\theta \theta}=r^2$ and,
\begin{equation}
g_{tt}=g_{rr}^{\;-1}=\left(1-\frac{2GM}{rc^2}\right)\;.
\end{equation}
Here $G$ is the gravitational constant and $M$ is the gravitational
mass of the object.  
Because of the assumption of the validity of the
equivalence principle, matter and photons follow geodesics in the
spacetime described by the metric~(\ref{eq:metric}).

We describe the motion of a massive particle in terms of its
4-velocity $u^\mu\equiv(u^t,u^r,u^\theta,u^\phi)$.  We choose our
coordinate system so that the orbit of the particles we study will lie
on the equatorial plane, i.e., we will set $\sin\theta=1$ and
$u^\theta=0$. The requirement $u_\mu u^\mu=-1$ for the 4-velocity of a
massive particle leads to the constraint
\begin{equation}
g_{tt}\left(u^t\right)^2+g_{rr}\left(u^r\right)^2+r^2\left(u^\phi\right)^2=-1\;.
\label{eq:vparticle}
\end{equation}

For a particle in a circular orbit at coordinate radius $r_e$, we
require that the radial component of its 4-velocity is zero, which
makes equation~(\ref{eq:vparticle}) a constraint on $u^t$ and $u^\phi$
This constraint is also true for the turning points in an elliptical
orbit. What sets a circular orbit apart is the fact that all points in
the trajectory are turning points, i.e., that
\begin{equation}
\frac{du^r}{dr}=0\;,
\end{equation}
which specifies the $u^t$ component of the 4-velocity uniquely
\begin{equation}
g_{tt} \left(u^t\right)^2\left(1-\frac{1}{2}\frac{d \ln |g_{tt}|}{d \ln r}\right) = -1\;.
\label{eq:lcirc}
\end{equation}
We can use expression~(\ref{eq:vparticle}) to solve for the $u^{\phi}$
component of the 4-velocity.  Whatever our theory of gravity is, we
expect that the outer regions of the galaxies are in the weak field,
so we can find an approximation to the desired accuracy by expanding
our metric away from the flat solution
\begin{equation}
g_{tt} = -1 + \epsilon^2 g_{tt}^{(2)}+\mathcal{O}(\epsilon)^3  \;,
\label{eq:gttexp}
\end{equation}
where we have introduced $\epsilon$ merely as a dummy parameter that
allows us to keep track of the expansion order. We denote the leading
order correction to the metric element as $\epsilon^2$, 
because
we are counting orders in terms of the expansion of the velocity,
which will be proportional to the square root of $g_{tt}^{(2)}$.
Using relations~(\ref{eq:vparticle})---(\ref{eq:gttexp}), we obtain
that the components of the 4-velocity of a particle in a circular
orbit are
\begin{eqnarray}
u^t &=& 1 +\epsilon^2\left[\frac{1}{2}\left(g_{tt}^{(2)}-\frac{1}{2}\frac{d g_{tt}^{(2)}}{d \ln r}\right)\right]+\mathcal{O}(\epsilon)^3\;,
\label{eq:ut}\\
u^r&=&0\;,\\
u^\theta&=&0\;,\\
u^\phi &=& \epsilon\left[ \frac{1}{r}\sqrt{-\frac{1}{2}\frac{d g_{tt}^{(2)}}{d \ln r}}\right]+\mathcal{O}(\epsilon)^3\;.
\label{eq:uphi}
\end{eqnarray}
We emphasize that the components of the 4-velocity of the particle in
circular orbit depend only on the value and the local radial
derivative of the $g_{tt}$ element of the metric.

For comparison, in a Schwarzschild metric, the non-zero components of
the 4-velocity of a particle in a circular orbit are given by
\begin{eqnarray}
u^t&\simeq & 1+3\frac{GM}{r}\\
u^\phi&\simeq & \frac{1}{r} \left(\frac{GM}{r}\right)^{1/2}\;.
\end{eqnarray}

\subsection{The Redshift of Photons}

Our next goal is to calculate the trajectories and energy shifts of
photons as they propagate from their origin in the galaxy to a distant
observer. For a photon with 4-momentum $k^\mu=(k^t, k^r, k^\theta,
k^\phi)$, there are conservation laws that arise from the two Killing
vectors $\xi^\mu=(1,0,0,0)$ and $\eta^\mu=(0,0,0,1)$ of the spacetime,
namely the conservation of energy
\begin{equation}
\varepsilon_{\rm p}\equiv-g_{\mu \nu}k^{\mu}\xi^{\nu}=-g_{tt}k^{t}
\end{equation}
and of angular momentum
\begin{equation}
l_{\rm p}\equiv g_{\mu \nu}k^{\mu}\eta^{\nu}=r^2 k^\phi\;.
\end{equation}
In its trajectory, the photon experiences an overall redshift and
Doppler shift, which is given by
\begin{equation}
1+z \equiv \frac{\nu_{\rm e}}{\nu_{\rm obs}} = 
\frac{g_{\mu \nu}(r_{\rm e})u_{\rm e}^{\mu} k_{\rm e}^{\nu}}
{g_{\mu \nu}(r_{\rm obs})u_{\rm obs}^{\mu} k_{\rm obs}^{\nu}}\;,
\label{eq:red_1}
\end{equation}
where the subscripts ``e'' and ``obs'' refer to the emitter and
the observer, respectively. 

Because of our assumption of asymptotic flatness, at the location of
the observer the spacetime is Minkowski and, therefore, $g_{tt}(r_{\rm
  obs}) \rightarrow -1$.  Moreover, because we are considering a
static observer, its 4-velocity is $u_{\rm obs}^{\mu}= (1,0,0,0)$, and
the denominator of the fraction in equation~(\ref{eq:red_1}) is equal
to $-\varepsilon_{\rm p}$. The 4-velocity of the emitting particle is
given by relations~(\ref{eq:ut})-(\ref{eq:uphi}), which after
inserting into equation~(\ref{eq:red_1}) leads to
\begin{equation}
z = - \frac{\epsilon}{r}\sqrt{-\frac{1}{2}\frac{d g_{tt}^{(2)}}{d \ln r}} \frac{l_{\rm p}}{\varepsilon_{\rm p}}+\frac{\epsilon^2}{2}\left(g_{tt}^{(2)}-\frac{1}{2}\frac{d g_{tt}^{(2)}}{d \ln r}\right)+\mathcal{O}(\epsilon)^3\;.
\label{eq:red}
\end{equation}
In this expression, we have dropped the subscripts `e' and `obs' for the
emitter and the observer. It is implicitly understood, however, that
the redshift $z$ is measured at the location of the observer, whereas
all the quantities in the right-hand side of the expression are
evaluated at the location of the emitter. This expression is identical at this
order to similar calculations based on different assumptions~\cite{rot}.

In order to calculate the quantity $l_{\rm p}/\varepsilon_{\rm p}$, we
first discuss the orientation and geometry of the galaxy, the
observer, and the photon trajectories in flat spacetime. All
corrections due to lensing appear as factors of at least order
$\mathcal{O}(\epsilon^3)$, which is beyond the order we are
considering here.

\begin{figure}
\includegraphics[width=10cm]{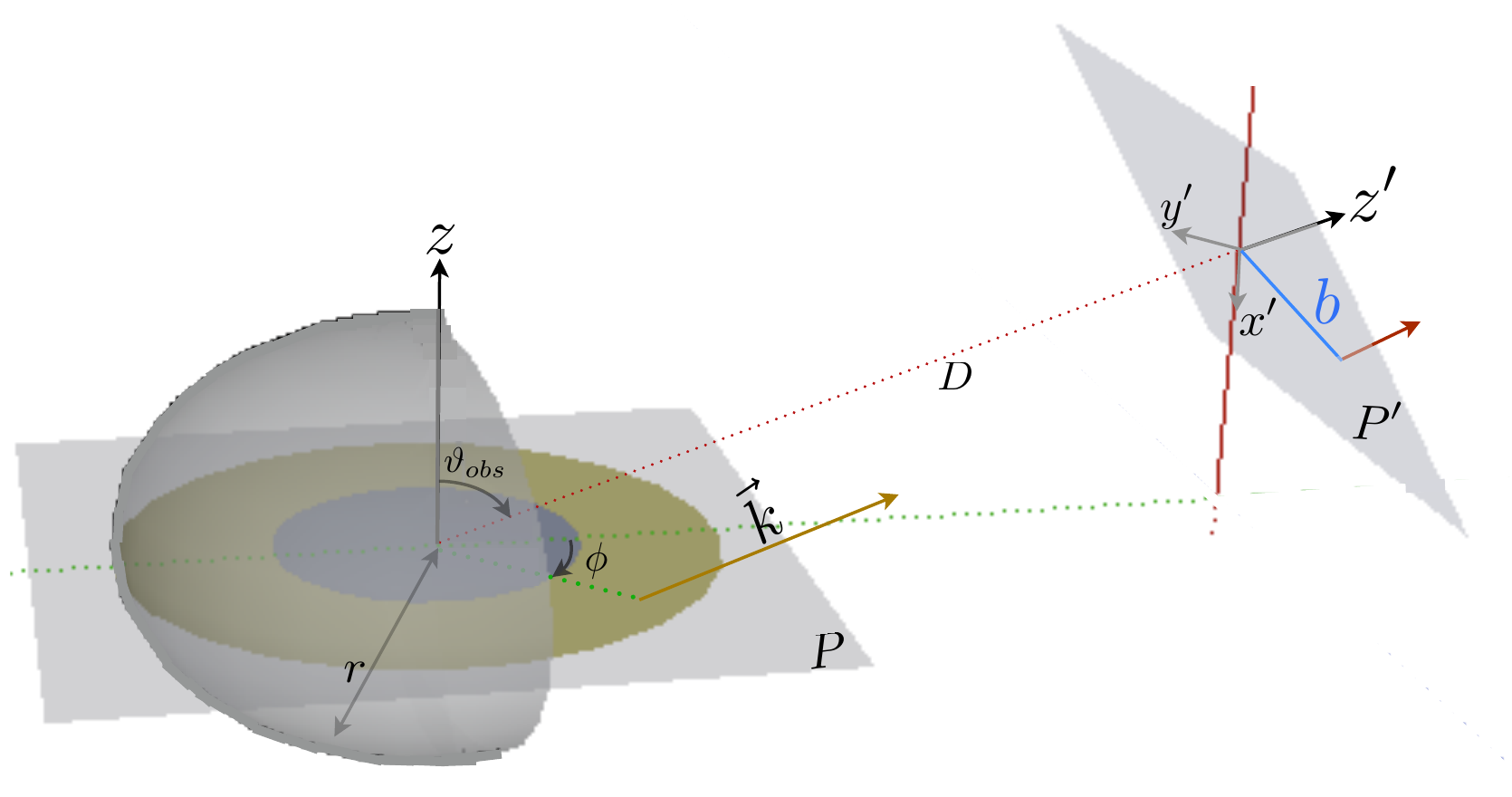}
\caption{A geometric representation of the galactic disk (grey ) plane
  $P$. Photons with momentum $\vec{k}$ are emitted within a ring (dark
  yellow) about the galactic center and arrive to the observer's image
  plane $P'$ with impact parameter $b$. The image plane is at some
  distance $D$ from the galactic plane, at an angle $\vartheta_{\rm
    obs}$ with respect to the axis of symmetry.  }
\label{fig}
\end{figure}

\subsection{The Trajectories of Photons}

We first set a coordinate system (see Fig.~\ref{fig}) with its origin
at the center of the galaxy and oriented in such a way that the orbits
of the emitting particles lie on the $x-y$ plane. The detector of the observer 
defines a second plane
(the image plane) at some great distance $D$ and at an angle
$\vartheta_{\rm obs}$ with respect to the direction of the angular
momentum of the galaxy $z$.
We use the axisymmetry of the
galaxy to choose the orientation of the $x-y$ axis so that the center
of the image plane of the distant observer lies on the $y-z$ plane. 

We then set a new coordinate system (indicated by primed
quantities) by rotating the original coordinate system around the
$x-$axis by angle $\vartheta_{\rm obs}$. The $x^\prime-y^\prime$ plane
of the new coordinate system is parallel to the image plane and the
unit vector to the image plane is parallel to the $z^\prime$ axis
\begin{equation}
\hat{z}' = \pmatrix{0 \cr \sin \vartheta_{\rm obs} \cr \cos \vartheta_{\rm obs}}\;.
\end{equation}

Coordinates on the image plane and coordinates on the galaxy plane are
related via the rotation
\begin{equation}
\pmatrix{x' \cr y' \cr z'} = \pmatrix{1 & 0 & 0 \cr 0 & \cos
  \vartheta_{\rm obs} & -\sin \vartheta_{\rm obs} \cr 0 & \sin
  \vartheta_{\rm obs} & \cos \vartheta_{\rm obs}} \pmatrix{x \cr y \cr z}\;.
\end{equation}
For an emitter in a circular orbit on the galactic plane at a radius
$r_e$ and at an azimuth $\phi_e$ with respect to the $x-$axis, this
relation becomes
\begin{equation}
\pmatrix{x' \cr y' \cr z'} = \pmatrix{1 & 0 & 0 \cr 0 & \cos
  \vartheta_{\rm obs} & -\sin \vartheta_{\rm obs} \cr 0 & \sin
  \vartheta_{\rm obs} & \cos \vartheta_{\rm obs}} \pmatrix{r_e \cos \phi_e
  \cr r_e \sin \phi_e \cr 0} = \pmatrix{r_e \cos \phi_e \cr r_e \cos
  \vartheta_{\rm obs} \sin \phi_e \cr r_e \sin \vartheta_{\rm obs} \sin
  \phi_e}
\label{eq:transf}
\end{equation}
If we do not consider the bending of the trajectory of a photon due to
gravitatonal lensing, then the $x^\prime$ and $y^\prime$ coordinates
calculated with the last relation will correspond to the location on
the image plane where the photon emitted by the orbiting object will
be detected. The impact parameter of that photon will, therefore, be
equal to
\begin{equation}
b = \left(x^{\prime 2}+y^{\prime 2}\right)^{1/2}
=r_c \sqrt{1 - \sin^2 \vartheta_{\rm obs}  \sin^2 \phi_e}\;.
\label{eq:imp}
\end{equation}

For a photon with wave vector $\vec{k}$ to intersect the image plane
at a right angle and with an impact parameter $b$ we require $\hat{k}
\parallel \hat{z}'$, so $\vec{k} = \left| k \right| \hat{z}'= \left| k
\right| \left( \sin \vartheta_{\rm o} \hat{y}+ \cos \vartheta_{\rm o}
\hat{z} \right)$. In spherical polar coordinates, for a photon emitted
in the plane $\theta = \frac{\pi}{2}$, the transformation from
cartesian to polar is straightforward and gives
\begin{equation}
\vec{k}= \left| k \right| \left( \sin \vartheta_{\rm obs} \sin \phi_e \;\hat{r} + \cos \vartheta_{\rm obs} \;\hat{\theta}+\sin \vartheta_{\rm obs} \cos \phi_e \;\hat{\phi} \right)\;.
\end{equation}
The null property of the photon 4-momentum requires 
\begin{equation}
\left| k \right| = \varepsilon_{\rm p} +\mathcal{O}(\epsilon^2)\;
\end{equation}
which leads to
\begin{equation}
\frac{l_{\rm p}}{\varepsilon_{\rm p}}= r_e\cos\phi_e +\mathcal{O}(\epsilon^2)\;.
\label{lovere}
\end{equation}

\subsection{Relativistic Redshifts Due to Galactic Rotation}

We are now in position to calculate the Doppler shift for an
axisymmetric potential and examine the properties particular to our
study of galaxies. From equation~(\ref{eq:red}) and~(\ref{lovere}) we
write
\begin{equation}
z = \epsilon z_1 \sin\vartheta_{\rm obs}\cos\phi_e+\epsilon^2 z_2+ \ldots\;,
\label{eq:red_exp}
\end{equation}
where
\begin{eqnarray}
z_1 &=& 
-\sqrt{-\frac{1}{2}\frac{d g_{tt}^{(2)}}{d \ln r}} \;,
\label{eq:z1}\\
z_2 &=& \frac{1}{2}\left(g_{tt}^{(2)}-\frac{1}{2}\frac{d g_{tt}^{(2)}}{d \ln r}\right)\;.
\label{eq:z2}
\end{eqnarray}
Combining the first two orders, we find
\begin{equation}
z_2=\frac{1}{2}\left(z_1^2+g_{tt}^{(2)}\right)\;.
\label{eq:z2_gen}
\end{equation}
This last relation expresses simply the fact that the second-order
energy shift has two distinct contributions: one from the second order
special-relativistic Doppler shift (captured by the first term in
the above sum) and one from the gravitational redshift (captured by
the second term in the above sum).

In a Schwarzschild spacetime, the expressions for the Doppler shift
plus redshift to all orders become
\begin{eqnarray}
z_1&=& -\left[\frac{GM}{r_{\rm c}c^2}\right]^{1/2}\\
z_2&=& \frac{3GM}{2r_{\rm c}c^2} \;,
\end{eqnarray}
where we introduced appropriate powers of the speed of light $c$ for
completeness.  Because the Schwarzschild spacetime has a single scale,
the relation between the first- and second-order terms is the
quadratic
\begin{equation}
z_2=\frac{3}{2}z_1^2
\label{eq:s_z2}
\end{equation}
and second-order effects are always suppressed compared to the first-order
effects.

Observations~\cite{sr01} suggest that, to leading order, the velocity profiles of
galaxies are very nearly flat over the radii of interest, i.e., 
\begin{equation}
z_1=-\left(\frac{u_0}{c}\right)\left(\frac{r_0}{r}\right)^{\alpha}\;,
\label{eq:gal_prof}
\end{equation}
where $u_0$ is the inferred, nearly constant rotational velocity,
at a characteristic length scale $r_0$ and we have introduced the
small parameter $\alpha\simeq 0$ to describe weak deviations from a
constant velocity profile. Such a rotation law implies that the
spacetime of the galaxy is described, to leading order, by
\begin{equation}
g_{tt}^{(2)} = \left(\frac{u_0}{c}\right)^2 \left\{
\begin{array}{ll}
\frac{1}{\alpha}
\left(\frac{r_0}{r}\right)^{2\alpha}&\;,\quad {\rm if}~\alpha > 0\\
\ln\left(\frac{r_{\rm out}}{r}\right)&\;,\quad {\rm if}~\alpha=0
\end{array}
\right.\;.
\label{eq:ord1}
\end{equation}
In this last expression for the special case $\alpha=0$, we have
introduced as an integration constant the radius $r_{\rm out}\gg r$ at
which the correction term $g_{tt}^{(2)}$ drops rapidly to zero. If a
dark matter halo with a density profile $\rho\sim r^{-2}$ is
responsible for the flat rotation curve of a galaxy, then $r_{\rm
  out}$ is the outer cut-off of the halo, which is necessary for the
total mass of the halo to be finite.  The exact value of this constant
does not affect the calculation of the redshift to first order.

Using equation~(\ref{eq:ord1}) for the $tt-$element of the metric, we 
now calculate the next order correction to the redshift as
\begin{equation}
z_2 = 
\left(\frac{u_0}{c}\right)^2 \left\{
\begin{array}{ll}
\frac{1+\alpha}{2\alpha}
\left(\frac{r}{r_0}\right)^{-2\alpha}&\;,\quad {\rm if}~\alpha< 0\\
\frac{1}{2}-\ln \sqrt{\frac{r_{\rm out}}{r}}&\;,\quad {\rm if}~\alpha=0
\end{array}
\right.\;.
\label{eq:ord2}
\end{equation}
For the case of a perfectly flat rotation curve (i.e., $\alpha=0$),
the sign and magnitude of the second-order wavelength shift $z_2$
depends explicitly on the cut-off radius $r_{\rm out}$.  To avoid the
additional complications introduced by the presence of the cut-off
radius, hereafter, we will assume $\alpha\ne 0$ and not discuss any
longer the singular case of a perfectly flat rotation
curve. Surprisingly, the second-order effects can become significant
for sufficiently flat rotation curves (i.e., when $\alpha\approx
0$). Formally speaking, our expansion is valid only for $\alpha \gtrsim
u_0/c$.

Note that the relation between the first- and second-order
effects for a galaxy with a nearly flat rotation curve is
\begin{equation}
z_2=\frac{1+\alpha}{2\alpha}z_1^2\simeq \frac{1}{2\alpha}z_1^2\;.
\label{eq:g_z2}
\end{equation}

Equations~(\ref{eq:ut}), (\ref{eq:uphi}), (\ref{eq:ord1}), and
(\ref{eq:ord2}) represent the main result of the last two sections;
that both the velocity profile of matter and the gravitational and
Doppler redshifts experienced by photons depend only on the same two
local properties of the metric at the place of emission: the value of
its $tt-$element and its radial derivative. Neither of the two
quantities depend on the field equations of the theory of gravity,
i.e., of the equation that determines the metric elements given a
distribution of matter.

\section{Doppler and Gravitational Corrections to the Line Profiles}

We now examine the implications of the results derived in the previous
section for the atomic line profiles detected from galaxies. The flux
an observer detects at a great distance $D$ is proportional to the
integral over the image plane of the specific intensity of rays that
arrive perpendicular to the image plane, i.e.,
\begin{equation}
F_\varepsilon(\varepsilon) =\frac{1}{D^2}\int dx' \int dy'\; 
I_\varepsilon(\varepsilon)\;.x
\end{equation}
Using the system of equations~(\ref{eq:transf}) we 
convert this
integral into one over coordinates in the galactic
plane
\begin{equation}
F_\varepsilon(\varepsilon) =\frac{\cos\vartheta_{\rm obs}}{D^2}\int 
r_{\rm e} dr_{\rm e} 
\int d\phi_{\rm e} I_\varepsilon(\varepsilon_{\rm e},r_{\rm e},\phi_{\rm e})
\left(\frac{\varepsilon}{\varepsilon_{\rm e}}\right)^3\;,
\end{equation}
where we have used the invariance of the quantity
$I_\varepsilon/\varepsilon^3$ to relate the intensity at arrival to
that at emission by
\begin{equation}
I_\varepsilon(\varepsilon) = I_{\varepsilon_{\rm e}}(\varepsilon_{\rm e})
\left(\frac{\varepsilon}{\varepsilon_{\rm e}}\right)^3\;.
\end{equation}
The observed energy $\varepsilon$ and the emitted energy $\varepsilon_{\rm e}$
are related by the redshift relations derived in the previous section
\begin{equation}
\frac{\varepsilon_{\rm e}}{\varepsilon}=1+z(r_{\rm e},\phi_{\rm e})
= 1+z_1 (r_{\rm e})\sin\vartheta_{\rm obs}\cos\phi_{\rm e}+z_2(r_{\rm e}) \;,
\end{equation}
where we have explicitly denoted the dependence of the redshift experienced
by each photon on the location of its emission.

We now assume that the emission at the local Lorentz frame is
mono-energetic, at a rest-frame energy $\varepsilon_0$. In other
words, we assume that
\begin{equation}
I_{\varepsilon_e}(\varepsilon_e,r_e,\phi_e)  
=\mathcal{I}(r_{\rm e},\phi_{\rm e})\delta\left[\varepsilon_e(r_e,\phi_e)-\varepsilon_0\right]
=\mathcal{I}(r_{\rm e}, \phi_{\rm e})
\delta\left\{\varepsilon\left[1+z(r_e,\phi_e)\right]-
\varepsilon_0\right\}\;.
\end{equation}
The flux integral, therefore, becomes
\begin{equation}
F_{\varepsilon}(\varepsilon)= \frac{\cos \vartheta_{\rm obs}}{D^2} 
\int dr_{\rm e} r_{\rm e}
\int d\phi_e  \mathcal{I}(r_{\rm e}, \phi_{\rm e})\delta\left\{\varepsilon-\varepsilon_0
\left[1+z(r_e,\phi_e)\right]^{-1}\right\}
\left[1+z(r_{\rm e},\phi_{\rm e})\right]^{-3}\;,
\end{equation}
where we have made a change of variables in the $\delta$-function to
reflect the fact that the right-hand side of this equation is a flux
density in the observed energy $\varepsilon$.  We use the
$\delta$-function to evaluate the integral over $\phi_{\rm e}$ using
the relation
\begin{equation}
\delta\left[g(\phi)\right]=\delta(\phi)\sum_i \left\vert
\frac{dg}{d\phi}\right\vert^{-1}_{\phi_i}\;,
\end{equation}
where $\phi_i$ is each solution to the equation $g(\phi_i)=0$ or, in
our case,
\begin{equation}
g(\phi_i)=0\Rightarrow \varepsilon-\varepsilon_0
\left[1+z(r_{\rm e},\phi_i)\right]^{-1}= 0\;.
\end{equation}
Using the expression~(\ref{eq:red_exp}) for the redshift,
we obtain
\begin{equation}
\cos\phi_i=\frac{1}{z_1\sin\vartheta_{\rm obs}}
\left(\frac{\varepsilon_0}{\varepsilon}-1-z_2\right)\;,
\end{equation}
which leads to two solutions for the angle $\phi_i$ with opposite
signs.  Requiring that $\vert\cos\phi_i\vert\le 1$ allows us to place
limits on the range of photon energies that contribute to the line as
\begin{equation}
\left(\frac{\varepsilon_\pm}{\varepsilon_0}\right)
=\frac{1}{1\pm z_1\sin\vartheta_{\rm obs}+z_2}\;.
\label{eq:eminmax}
\end{equation}

In evaluating the integral, we also need the derivative
\begin{equation}
\left\vert\frac{dz}{d\phi_{\rm e}}\right\vert_{\phi_{\rm e}=\phi_i}=
\left[z_1^2\sin^2\vartheta_{\rm obs}-
\left(\frac{\varepsilon_0-\varepsilon}{\varepsilon}\right)^2+
2\left(\frac{\varepsilon_0-\varepsilon}{\varepsilon}\right)z_2
\right]^{1/2}\;.
\end{equation}
Using the above expressions, the flux becomes
\begin{eqnarray}
F_{\varepsilon}(\varepsilon)&=& \frac{\cos \vartheta_{\rm obs}}{D^2} 
\int dr_{\rm e} r_{\rm e}\sum_i
\mathcal{I}(r_{\rm e},\phi_{i})\left[1+z(r_{\rm e},\phi_i)\right]^{-3}
\left\vert\frac{dg}{d\phi_{\rm e}}\right\vert_{\phi_e=\phi_i}^{-1}
\nonumber \\
&=& 2\frac{\cos \vartheta_{\rm obs}}{D^2} 
\int dr_{\rm e} r_{\rm e}
\left[1+z(r_{\rm e},\phi_i)\right]^{-1}
\left\vert\frac{dz}{d\phi_e}\right\vert_{\phi_e=\phi_i}^{-1}\left[\frac{\mathcal{I}(r_{\rm e},\phi_i)+\mathcal{I}(r_{\rm e},-\phi_i)}{2}\right]\nonumber\\
&=& 2\frac{\cos \vartheta_{\rm obs}}{D^2} 
\left(\frac{\varepsilon}{\varepsilon_0}\right)
\int dr_{\rm e} r_{\rm e}
\left\vert\frac{dz}{d\phi_e}\right\vert_{\phi_e=\phi_i}^{-1}\left[\frac{\mathcal{I}(r_{\rm e},\phi_i)+\mathcal{I}(r_{\rm e},-\phi_i)}{2}\right]\nonumber\\
&\simeq& 2\frac{\cos \vartheta_{\rm obs}}{D^2} 
\left(\frac{\varepsilon}{\varepsilon_0}\right)
\int dr_{\rm e} r_{\rm e}
\left[z_1^2\sin^2\vartheta_{\rm obs}-
\left(\frac{\varepsilon_0-\varepsilon}{\varepsilon}\right)^2+
2\left(\frac{\varepsilon_0-\varepsilon}{\varepsilon}\right)z_2
\right]^{-1/2}\left[\frac{\mathcal{I}(r_{\rm e},\phi_i)+\mathcal{I}(r_{\rm e},-\phi_i)}{2}\right]\;.
\label{eq:flux}
\end{eqnarray}

To explore the properties of the second-order corrections to the line
profiles, we make for now the simplifying assumption that the emission
comes from a single annulus in the galactic disk with intensity that
is independent of azimuth, at a radius $r_0$ and with a width $\delta
r$. In this case, the line profile becomes
\begin{equation}
F_\varepsilon(\varepsilon)=
2\cos \vartheta_{\rm obs} \mathcal{I}(r_0)\left(\frac{r_0 \delta r}{D^2}\right)
\left(\frac{\varepsilon}{\varepsilon_0}\right)
\left[z_1^2\sin^2\vartheta_{\rm obs}-
\left(\frac{\varepsilon_0-\varepsilon}{\varepsilon}\right)^2+
2\left(\frac{\varepsilon_0-\varepsilon}{\varepsilon}\right)z_2
\right]^{-1/2}\;,
\end{equation}
which we express in terms of the dimensionless quantity
\begin{eqnarray}
\mathcal{F}_\varepsilon&\equiv&\frac{F_\varepsilon(\varepsilon)}
{2\cos \vartheta_{\rm obs} \mathcal{I}(r_0)}
\left(\frac{r_0 \delta r}{D^2}\right)^{-1}\nonumber\\
&=&
\left(\frac{\varepsilon}{\varepsilon_0}\right)
\left[z_1^2\sin^2\vartheta_{\rm obs}-
\left(\frac{\varepsilon_0-\varepsilon}{\varepsilon}\right)^2+
2\left(\frac{\varepsilon_0-\varepsilon}{\varepsilon}\right)z_2
\right]^{-1/2}\;,
\end{eqnarray}

The minimum and maximum photon energies for which the flux is non-zero
are given by equation~(\ref{eq:eminmax}). Because of the second-order
Doppler shifts and the gravitational redshift, the center of the
broadened profile is displaced from the rest energy of the line by an
amount equal to
\begin{equation}
\bar{\varepsilon}=\frac{1}{2}\left(\varepsilon_-+\varepsilon_+\right)=
\varepsilon_0\left(1+z_1^2\sin^2\vartheta_{\rm obs}-z_2\right)\;.
\end{equation}
Moreover, the amplitude of the red wing of the line is smaller than
the amplitude of the blue wing.

\begin{figure}
\includegraphics[width=6cm]{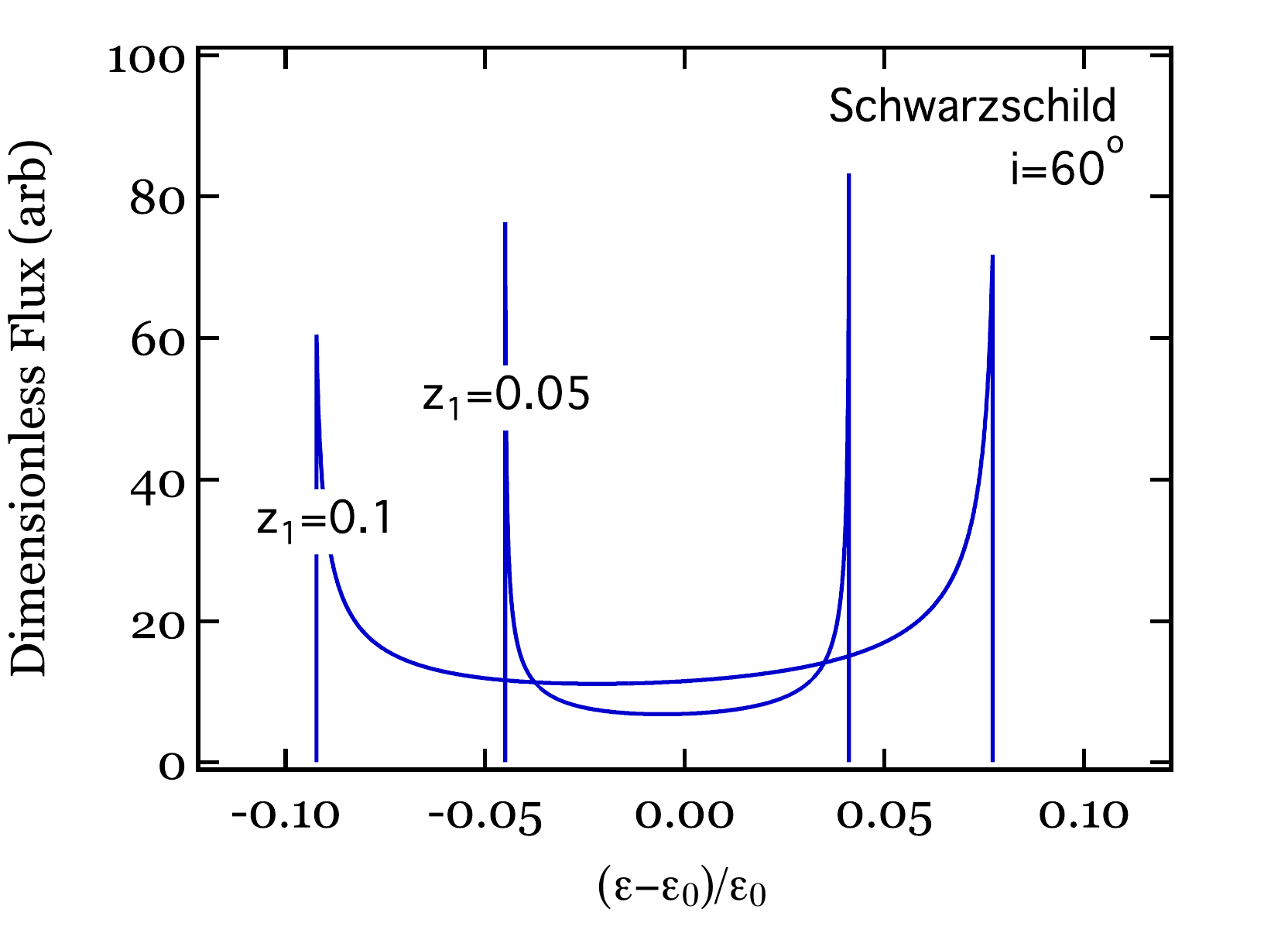}
\includegraphics[width=6cm]{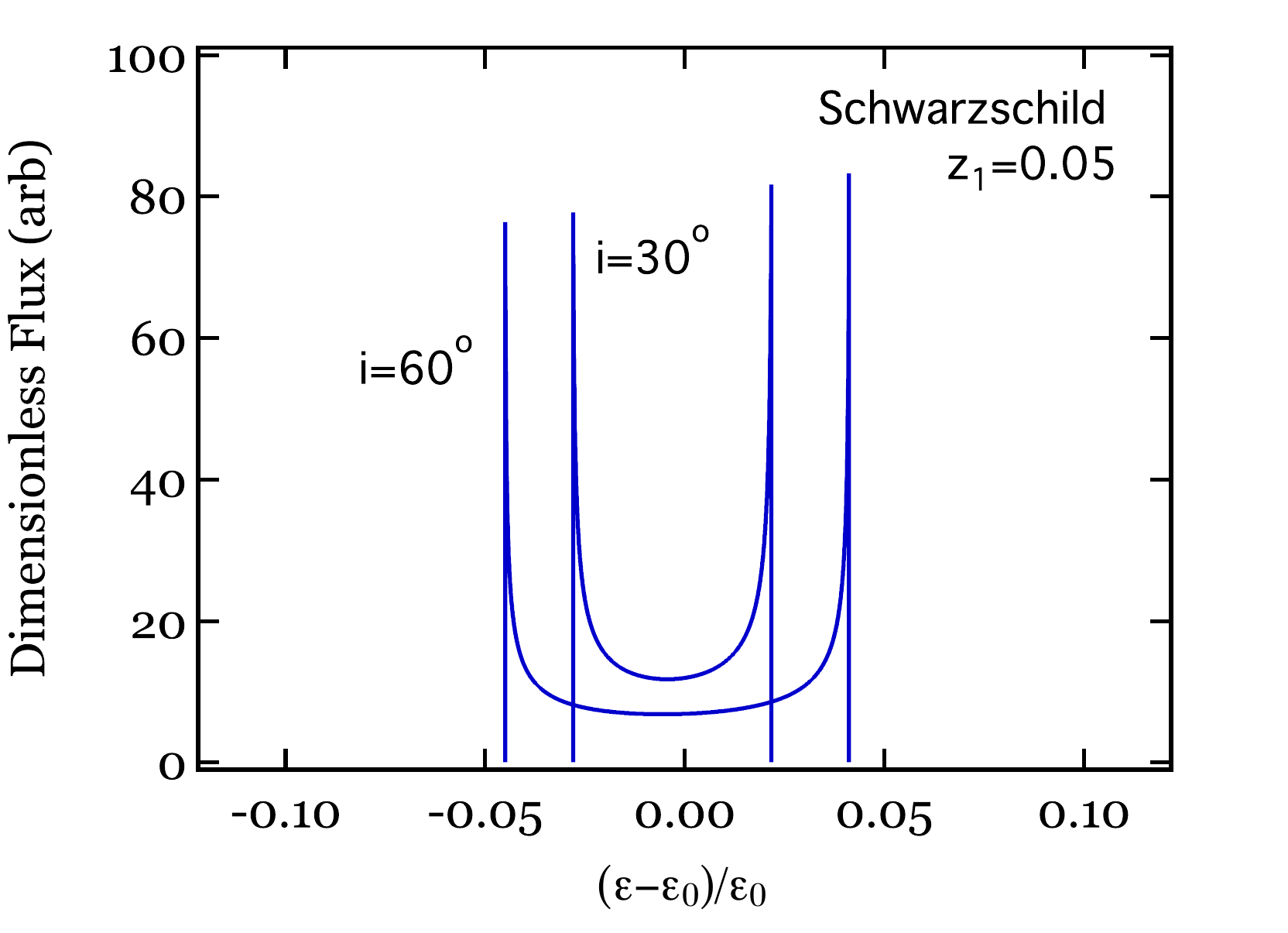}
\caption{Profiles of broadened atomic lines that originate in an annulus
in a Schwarzschild spacetime for different values of the non-relativistic
Doppler shift $z_1$ and inclinations of the observer. The overall effect of
the relativistic Doppler shift and of the gravitational redshift is to make
the profile asymmetric and to shift it towards lower energies.}
\label{fig:scw}
\end{figure}

\begin{figure}
\includegraphics[width=7cm]{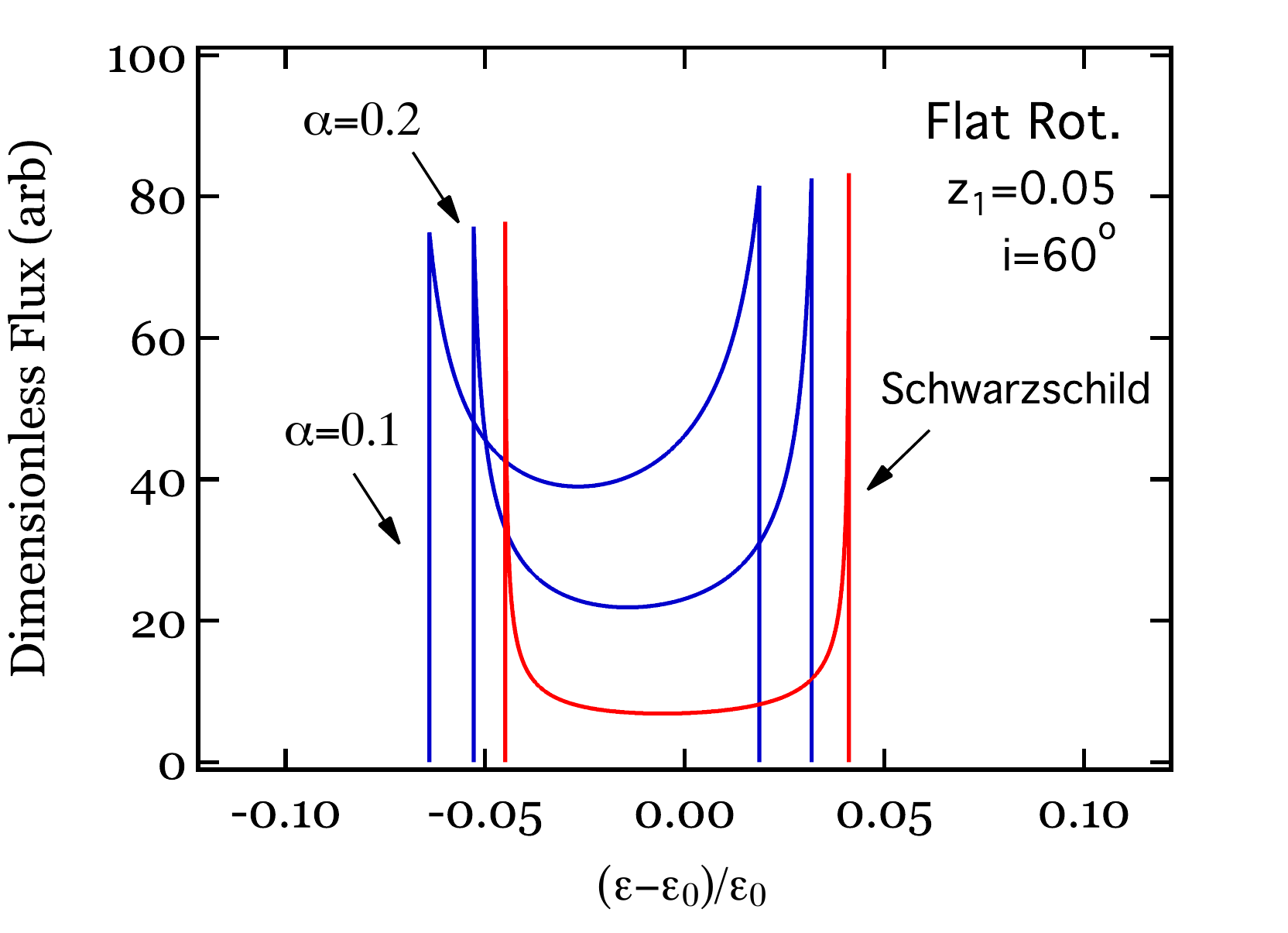}
\caption{Profiles of broadened atomic lines that originate in an
  annulus in a galactic disk plotted for different values of the 
  parameter $\alpha$, which measures the degree of flatness of the
  galactic rotation curve. The red line shows, for comparison, the
  line in a Schwarzschild spacetime with the same amount of
  non-relativistic Doppler shift. The spacetime that corresponds to a
  flat rotation curve leads to a larger overall redshift as well as to
  less pronounced blue and red wings for the line. Note that the
  parameter $z_1$ used for this plot is unphysically large for a
  galaxy and was chosen here in order to make the effect visible.}
\label{fig:gal}
\end{figure}

Figures~(\ref{fig:scw}) and (\ref{fig:gal}) show the dimensionless
line profiles from narrow annuli in a Schwarzschild spacetime and in
the spacetime of a galaxy with a nearly flat rotation curve, for
different values of the parameters $z_1$ and $\alpha$ (see
equation~[\ref{eq:s_z2}] and [\ref{eq:g_z2}]) and for different inclinations
of the observer. In the case of a galaxy with a nearly flat rotation
curve, the broadened line is gravitationally redshifted by an amount
that is larger compared to the equivalent Schwarzschild case. If the
flat rotation curve is a result of a dark matter halo, the additional
redshift occurs because the dark matter halo has a density profile
$\rho\simeq r^{-2}$ and a large amount of mass (and hence of
gravitational redshift) exists outside the location of the annulus. If
this second-order effect is not corrected for, it will be assigned to
the overall recession velocity of the galaxy $V_{\rm g}$ that is due
to its peculiar motion and to the Hubble flow.  The error is,
nevertheless, small as it is of order
\begin{eqnarray}
\left\vert\frac{\delta V_{\rm g}}{V_{\rm g}}\right\vert&\simeq& 
\left\vert\frac{\bar{\varepsilon}-\varepsilon_0} {\varepsilon_0}\right\vert=
\left\vert\sin^2\vartheta_{\rm obs}-\frac{1+\alpha}{2\alpha}\right\vert
z_1^2\nonumber\\
   &\simeq&\frac{1}{2\alpha}
   \left(\frac{V_{\rm rot}}{c}\right)^2=
   \frac{1}{2}\left(\frac{V_{\rm rot}}{\alpha c}\right)
   \left(\frac{V_{\rm rot}}{c}\right)
   \le 
   \frac{1}{2}
   \left(\frac{V_{\rm rot}}{c}\right)\ll 1
\qquad \mbox{when}~~\alpha\ge\frac{V_{\rm rot}}{c}\;,
\end{eqnarray}
where $V_{\rm rot}$ is the inferred rotational velocity from the
non-relativistic Doppler shift.

\section{A Statistical Measure of the Relativistic Doppler Shift and of the 
Gravitational Redshift}

In the previous section, we calculated the profile of an atomic line
that originates in the equatorial plane of a galaxy in a general
metric theory of gravity. We showed that the non-relativistic Doppler
shift, the relativistic (i.e., second-order) Doppler shift, and the
gravitational redshift experienced by a photon from its origin to a
distant observer depend on the magnitude and local radial derivative
of the $tt$-element of the metric, $g_{tt}$, at its
origin. We derive here quantities that enable the measurement 
of these second-order corrections. 

We start by considering a single galaxy at a redshift $z_{\rm g}$,
which accounts for both the peculiar velocity of the galaxy as well as
the Hubble flow. We will assume that the rotational profile of the
galaxy, as inferred from non-relativistic Doppler shifts, is nearly
flat and described by equation~(\ref{eq:gal_prof}) from some inner
radius $r_0$ to an outer radius $\gg r_0$, i.e.,
\begin{equation}
z_1=-\left(\frac{u_0}{c}\right)\left(\frac{r}{r_0}\right)^{-\alpha}\;.
\end{equation}

The relativistic Doppler shift and the gravitational redshift
introduce an additional overall change in the energy of the photon
that we described by the quantity $z_2$ given, in general, by
equation~(\ref{eq:z2_gen}). If photons and particles follow geodesics
in the same spacetime, i.e., if the theory of gravity obeys the
equivalence principle, then the gravitational redshift can be
calculated using the same metric that determines the velocities of the
emitting hydrogen atoms. In this case, $z_2$ is related to $z_1$
according to equation~(\ref{eq:g_z2}). If, on the other hand, the
theory of gravity does not obey the equivalence principle, then the
amount of gravitational redshift experienced by each photon will not
have the same relation to the non-relativistic Doppler shifts. We parametrize
the possibility of the theory not obeying the equivalence principle by
the single constant $f$ and write, in general, the second-order energy
shift as
\begin{equation}
z_2=\frac{f+\alpha}{2\alpha}z_1^2\;.
\label{eq:g_z2_gen}
\end{equation}
For a gravity theory that obeys the equivalence principle, $f\equiv 1$.

The line profile measured by an observer at infinity will extend
between the two energies $\varepsilon_-$ and $\varepsilon_+$ dominated
by the largest rotational velocities and, therefore, (see
eq.~[\ref{eq:eminmax}])
\begin{equation}
\varepsilon_\pm=
\frac{\varepsilon_0}{1+z_{\rm g}}
\left[1\pm\left(\frac{u_0}{c}\right)\sin\vartheta_{\rm obs}
+\left(\frac{u_0}{c}\right)^2\left(\sin^2\vartheta_{\rm obs}-\frac{f+\alpha}{2\alpha}\right)
\right]\;.
\end{equation}
The width of the line is 
\begin{equation}
\Delta\varepsilon \equiv \varepsilon_+ - \varepsilon_-=\frac{\varepsilon_0}{1+z_{\rm g}}\left[2\left(\frac{u_0}{c}\right)\sin\vartheta_{\rm obs}\right]
\label{eq:width}
\end{equation}
and is determined by the non-relativistic Doppler shift. On the other
hand, the center of the line is
\begin{equation}
\bar{\varepsilon}\equiv\frac{1}{2}\left(\varepsilon_++\varepsilon_-\right)=
\frac{\varepsilon_0}{1+z_{\rm g}}\left[1+\left(\frac{u_0}{c}\right)^2\left(\sin^2\vartheta_{\rm obs}-\frac{f+\alpha}{2\alpha}\right)
\right]
\label{eq:eave}
\end{equation}
and deviates from $\varepsilon_0/(1+z_{\rm g})$ by second order
effects. These two gross properties of the line profile depend on four
parameters of the galaxy and on the gravitational theory: the redshift
of the galaxy $z_{\rm g}$, its rotational velocity $u_0$, the
inclination of the observer $\vartheta_{\rm obs}$, and the ratio
$f/\alpha$. In principle, we can perform an independent
measurement of the redshift $z_g$ of the galaxy using optical lines
from the galactic nucleus. For a sufficiently large sample of galaxies,
we may also use a statistical argument regarding the distribution of
inclinations. Hence we can use precise measurements of both
$\Delta\varepsilon$ and $\bar{\varepsilon}$ for a large number of
galaxies in order to measure statistically the parameter $f/\alpha$
and, hence, constrain deviations from the equivalence principle.

Because the above argument is of a statistical nature, it will be
difficult to determine the formal and, more importantly, the
systematic uncertainties of the result based solely on this single
type of measurement.  There is, however, an additional measureable
quantity that provides an independent measure of the same parameters
and can be used as a consistency check in case any deviations from the
equivalence principle are found.  The relativistic-Doppler shift and
the gravitational redshift not only introduce an additional energy
shift to the line, but also make it asymmetric, with the blue wing
appearing brighter than the red wing. As a result, we obtain an
independent third observable from each galaxy in the flux averaged
photon energy
\begin{equation}
 \langle\varepsilon\rangle \equiv \frac{\int d \varepsilon \;\;\varepsilon 
\;F_{\varepsilon}(\varepsilon)}
{\int d \varepsilon \; F_{\varepsilon}(\varepsilon)} \;.
\end{equation}
For a symmetric line,
$\langle\varepsilon\rangle=\bar{\varepsilon}$; any correction to this
equality will be due to the second-order relativistic effects.

For simplicity, we first perform the calculation of $\langle
\varepsilon \rangle$ in the frame of the galaxy and add the redshift
due to the Hubble flow and its peculiar velocity only in the final
result. In deriving the flux averaged photon energy, we need to
evaluate two integrals of the form (see eq.~[\ref{eq:flux}])
\begin{eqnarray}
E_n&=&\int d \varepsilon\;\varepsilon^nF_{\varepsilon}(\varepsilon)\nonumber\\
&=&2\frac{\cos \vartheta_{\rm obs}}{D^2} 
\int dr_{\rm e} r_{\rm e}
\int d\varepsilon\;\varepsilon^n
\left[1+z(r_{\rm e},\phi_i)\right]^{-1}
\left\vert\frac{dz}{d\phi_e}\right\vert_{\phi_e=\phi_i}^{-1}\left[\frac{\mathcal{I}(r_{\rm e},\phi_i)+\mathcal{I}(r_{\rm e},-\phi_i)}{2}\right]\;.
\end{eqnarray}
We write
\begin{equation}
\varepsilon=\varepsilon_0\frac{1}{1+z(r_{\rm e},\phi_i)}
\end{equation}
and make the change of variables from $\varepsilon$ to $\phi_i$ 
\begin{equation}
d\varepsilon = -\varepsilon_0\left[\frac{1}{1+z(r_{\rm e},\phi_i)}\right]^2
\left.\frac{dz}{d\phi_{\rm e}}\right\vert_{\phi_e=\phi_i}d\phi_i \;.
\end{equation}
As a result, the general expression for each integral becomes
\begin{eqnarray}
E_n&=&2\varepsilon_0^n\frac{\cos \vartheta_{\rm obs}}{D^2} 
\int dr_{\rm e} r_{\rm e}
\int_{0}^\pi d\phi_i
\left[1+z(r_{\rm e},\phi_i)\right]^{-3-n}\left[\frac{\mathcal{I}(r_{\rm e},\phi_i)+\mathcal{I}(r_{\rm e},-\phi_i)}{2}\right]\nonumber\\
&\simeq&\varepsilon_0^n\frac{\cos \vartheta_{\rm obs}}{D^2} 
\int dr_{\rm e} r_{\rm e}\int_{0}^{2 \pi} d\phi_i\mathcal{I}(r_{\rm e},\phi_i)
\left[1+z_1\sin\vartheta_{\rm obs}\cos\phi_i+z_2\right]^{-3-n}\nonumber\\
&\simeq&\varepsilon_0^n\frac{\cos \vartheta_{\rm obs}}{D^2} 
\int  r_{\rm e}dr_{\rm e}d\phi_i
\mathcal{I}(r_{\rm e},\phi_i) \left\{1-\left(3+n\right)\left[z_1\sin\vartheta_{\rm obs} \cos \phi_i +z_2
-\frac{(4+n)}{2}z_1^2\sin^2\vartheta_{\rm obs}\cos^2 \phi_i
\right]\right\}\;.
\end{eqnarray}
Inserting first the general relation~(\ref{eq:g_z2_gen}) between the
first- and second-order energy shifts, we obtain
\begin{eqnarray}
E_n = \varepsilon_0^n\frac{\cos \vartheta_{\rm obs}}{D^2}\left\{
\int r_e dr_e d\phi_i \mathcal{I}\left(r_e,\phi_i
\right)+\left(3+n\right)\left[\frac{\left(4+n\right)}{2}\sin^2\vartheta_{\rm obs}\int r_e dr_e d\phi_i
\mathcal{I}\left(r_e,\phi_i
\right)z_1^{\;2} \cos^2 \phi_i \right. \right. \nonumber\\
-\left.\left(\frac{f+\alpha}{2\alpha}\right)\int r_e dr_e d\phi_i
\mathcal{I}\left(r_e,\phi_i
\right)z_1^{\;2} -\sin \vartheta_{\rm obs}
\left.\int r_e dr_e d\phi_i
\mathcal{I}\left(r_e,\phi_i
\right)z_1 \cos \phi_i \right]\right\}\;.
\end{eqnarray}
We now write this expression more compactly by defining an intensity average of 
quantities involving $z_1$ and  $\cos \phi_i$ 
as follows
\begin{equation}
 \langle z_1^\alpha \cos^\beta \phi_i\rangle =\frac{\int r_e dr_e d\phi_i\;
\mathcal{I}\left(r_e,\phi_i
\right)z_1^\alpha \cos^\beta \phi_i }{\int r_e dr_e d\phi_i \;\mathcal{I}\left(r_e,\phi_i
\right)}
\end{equation}
The flux averaged photon energy is then given by
\begin{equation}
\langle \varepsilon\rangle=\frac{E_1}{E_0}=\frac{\varepsilon_0}{1+z_{\rm g}}
\left[1-\sin \vartheta_{\rm obs} \langle z_1\cos \phi \rangle+9\sin^2 \vartheta_{\rm obs}\langle z_1\cos \phi \rangle^2+4\sin^2\vartheta_{\rm obs}\langle z_1^{\;2}\cos ^2 \phi \rangle
-\frac{f+\alpha}{2\alpha}\langle z_1^{\;2}\rangle\right]
\;.
\label{eq:eang}
\end{equation}

The fluxed average photon energy in each galaxy depends strongly on
the degree of asymmetry in the emission. However, we can again make a
statistical measurement of the parameter $f/\alpha$ using a large
sample of galaxies. Indeed, because the angle $\phi_i$ is measured
with respect to the distant observer, its values for different
galaxies will not be correlated with the first-order redshift $z_1$
but will be randomly distributed. As a result, in a statistical
sense, equation~(\ref{eq:eang}), when averaged over all possible
azimuthal orientations of the observer, will be
\begin{equation}
\langle \varepsilon\rangle=\frac{E_1}{E_0}=\frac{\varepsilon_0}{1+z_{\rm g}}
\left[1+\left(\frac{u_0}{c}\right)^2\left(2\sin^2\vartheta_{\rm obs}
-\frac{f+\alpha}{2\alpha}\right)\right]
\label{eq:eang_stat}
\;,
\end{equation}
where we have also taken advantage of the slow variation of $z_1$
with radius $r_e$. When combined with equation~(\ref{eq:width}) for
the width of the line and averaged over all possible inclinations of
the observer $\theta_{\rm obs}$, the flux averaged energies of a large
sample of galaxies will also lead to an independent measurement of
the parameter $f/\alpha$ and hence to a test of the equivalence
principle.

\section{Discussion}

\begin{figure}
\includegraphics[width=6cm]{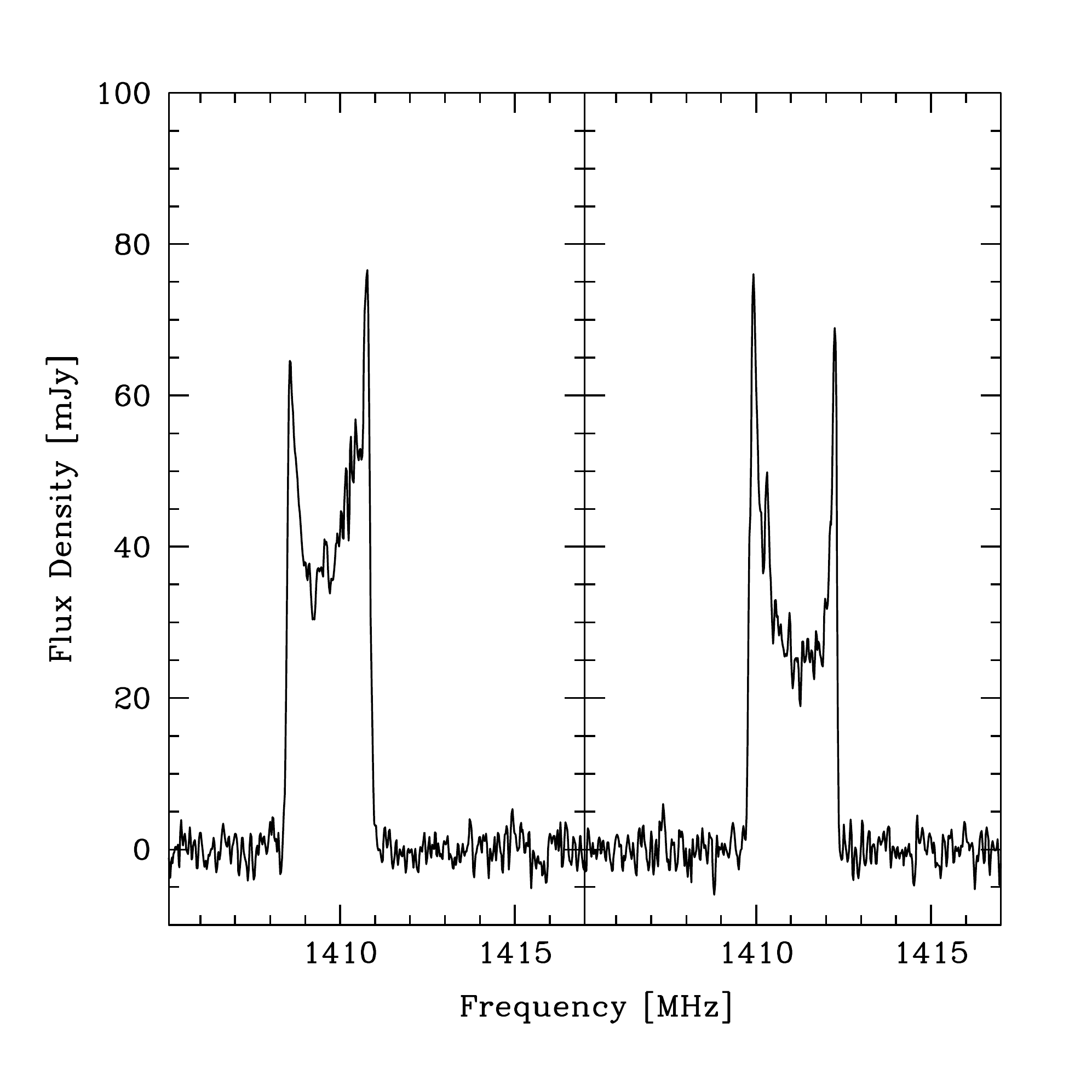}
\caption{ H{\small I} profiles of two massive galaxies ($V_{rot} >
  200$ km/sec) that are among those with the highest signal-to-noise
  ratio in the sample. The two profiles show different signs of the
  asymmetry, reflecting the ``noise" introduced in such measurements
  by the internal, structural asymmetries of the galaxies.}
\label{fig:hiprofiles}
\end{figure}

In this article, we calculated the second-order special relativistic and
gravitational effects on the rotationally broadened line profiles from
galaxies. We identified two measurable quantities 
(eqs.~[\ref{eq:eave}] and [\ref{eq:eang_stat}]),
which can be used in conjuction with the observed line 
widths~(eq.~[\ref{eq:width}]) to test the validity of the equivalence
principle at galactic scales. 

The level at which we can perform this test depends on the formal 
uncertainties in each measurement, as well as on the degree of 
azimuthal symmetry in the emission of each galaxy, which can mask
the asymmtry in the line profile due to relativistic effects. 
To estimate the magnitude of these two sources of uncertainty, we examine
some H{\small I} profiles provided as results of ongoing single beam
surveys~\cite{hicats}. 

As evident from Fig.~\ref{fig:hiprofiles}, line profiles for
individual galaxies, even when they are of high signal-to-noise and
sharply double peaked, can have morphologies quite different from the
azimuthally symmetric case of Fig.~\ref{fig:gal}. This variance is
because the H{\small I} is not necessarily smoothly distributed and
galaxies often are lopsided in the stellar or gaseous distribution
\cite{lopsided}. Therefore, even if the effect we are searching for is
large, which it is is not, the nature of this measurement must be
statistical.

We also estimate the internal uncertainties using an estimate of the
spectral noise from regions outside of the profile and 1000
simulations with noise added to the spectra. Because this noise is
superposed on the original spectrum, the noise of the simulations is
actually $\sqrt{2}$ larger, resulting in a slight overestimate of the
internal uncertainties. We find that these are always smaller than the
systematic errors due to intrinsic profile asymmetries.  

A precision measurement of the parameter $f$, which measures potential
violations of the equivalence principle, requires great care in
handling of the data as subtle biases can be introduced by careless
binning or scaling. While we argued earlier that we could average
over a suitably large distribution of inclinations, it is, in
principle, possible to obtain independent measurements of the
inclinations of individual galaxies which would further restrict the
range of the constraints. New surveys will be releasing many thousands
of galaxy spectra in the near future, which offers the possibility of
placing 
constraints on potential equivalence principle
violations on the scale of galaxies. We will discuss a detailed 
observational strategy for dealing with these issues in a forthcoming
article.

\acknowledgements

We thank Feryal \"Ozel for useful discussions. This work was supported
in part by the NSF CAREER award NSF~0746549 to DP and NSF AST-0907771 to DZ.

\bibliographystyle{apsrev}

\end{document}